\documentclass[letterpaper,showpacs,preprintnumbers,amsmath,amssymb,nofootinbib, twocolumn, superscriptaddress,notitlepage,prd]{revtex4-1}

\usepackage[utf8]{inputenc}

\usepackage{ulem}

\usepackage{graphicx}
\usepackage[usenames,dvipsnames]{color}
\usepackage[
colorlinks=true,       
linkcolor=red,          
citecolor=blue,        
filecolor=magenta,      
urlcolor=cyan  
]{hyperref}

\hypersetup{colorlinks,citecolor= blue,linkcolor= blue, urlcolor=blue}

\usepackage{amsmath}
\usepackage{bbm}
\usepackage{amsfonts}
\usepackage{amssymb}
\usepackage{latexsym}
\usepackage{graphicx}
\usepackage[english]{babel}
\usepackage{multirow}
\usepackage{float}
\usepackage{url}
\usepackage{slashed}
\usepackage{xcolor} 
\usepackage{array}
\usepackage{cases}

\newcommand{\be}{\begin{equation}}
\newcommand{\ee}{\end{equation}}
\newcommand{\ba}{\begin{array}}
\newcommand{\ea}{\end{array}}
\newcommand{\bea}{\begin{eqnarray}}
\newcommand{\eea}{\end{eqnarray}}

\newcommand{\eeccp}{$e^+ e^- \to \chi\bar \chi \pi^0$\,}

\newcommand{\GeV}{\,{\rm GeV}}

\usepackage{ulem,fancyvrb}
\usepackage{xcolor}

\begin{document}
\title{Can millicharge be probed at  future Super Tau Charm Facility via mono-$\pi^0$ searches?}

\author{Yu Zhang}
\email{dayu@hfut.edu.cn}
\affiliation{School of Physics, Hefei University of Technology, Hefei 230601, P.R.China}

\begin{abstract}
We propose a new channel to search for millicharged particles at the future Super Tau Charm Facility (STCF) via mono-$\pi^0$ signature. For the first time, we compute the mono-$\pi^0$ signal events at the future STCF due 
to millicharged particle production, as well as due to standard model 
irreducible/reducible backgrounds. 
By utilizing the assumed 20 ab$^{-1}$ luminosity for each running energy with $\sqrt{s}= 2$ GeV, 4 GeV and 7 GeV,
we derive the corresponding upper limits on millicharge, respectively.
Via  mono-$\pi^0$ searches  at the future STCF with $\sqrt{s}=2$ GeV, the upper limits on millicharge can be improved than ArgoNeuT when the mass of millicharged particle is less than about 500 MeV, but are not very competitive compared to a latest derivation from a past BEBC experiment and a new SENSEI experiment. Regardless, the mono-$\pi^0$ searches could be an important complement to investigate the invisible particles, such as dark matter.
\end{abstract}

\maketitle

\section{Introduction}
A vital direction of modern particle physics is the search for new particles beyond the Standard Model (SM),
which are very weakly coupled. 
Since there is no clear evidence on the existence of magnetic monopoles, 
which can quantize the electric charge  \cite{Dirac:1931kp},
theoretically, the charge
of particle can be arbitrary.
Normally we refer new particles beyond SM  with very small electric charge
as millicharged or minicharged particles (MCPs).
We employ the following 
interaction Lagrangian 
\be
{\cal L}_\text{int} = e \varepsilon A_\mu \bar \chi \gamma^\mu \chi, 
\label{eq:millicharge}
\ee
to parameterize the extremely weak coupling between a millicharged 
fermion and the SM photon,
where $\chi$ is the millicharged particle, 
$A_\mu$ is the SM photon, 
and $\varepsilon$ is the millicharge normalized to 
the magnitude of the electron charge. 
There are viable nature ways to introduce the millicharge.
Such as MCP  may be present in models with a kinetic mixing portal \cite{Holdom:1985ag, Holdom:1986eq, Foot:1991kb}, or in a Stueckelberg portal \cite{Kors:2004dx, Feldman:2007wj, Cheung:2007ut}.

In terrestrial accelerator experiments, many efforts have been made to look for MCPs.
Despite the negative results, the upper limit on millicharge can be obtained at collier experiments \cite{Davidson:1991si, Davidson:2000hf, CMS:2012xi, Haas:2014dda, Ball:2020dnx} and fixed-target experiments with a proton beam \cite{Magill:2018tbb, Harnik:2019zee, ArgoNeuT:2019ckq, Marocco:2020dqu, SENSEI:2023gie}  or with an electron beam \cite{Prinz:1998ua, Golowich:1986tj, Soper:2014ska, Berlin:2018bsc, Gninenko:2018ter, Chu:2018qrm, Anchordoqui:2021ghd}.  A number of 
future experiments with sensitivity to MCPs at accelerator facilities have been proposed \cite{Ball:2016zrp, Liu:2018jdi, Kelly:2018brz, Liu:2019ogn, Foroughi-Abari:2020qar, Liang:2019zkb, Afek:2020lek, Kim:2021eix, Gorbunov:2021jog, Carney:2021irt, Budker:2021quh, Kling:2022ykt, Gorbunov:2022dgw, Oscura:2023qch,Gorbunov:2022bzi,Kalliokoski:2023cgw}.

Thanks to their non-zero coupling with the SM photon, the MCPs can be produced in pairs by a virtual photon at particle colliders, such as $e^+e^-\to\chi\bar\chi$ at electron-positron colliders.
However, the produced MCPs are often undetectable in practice, because only a feeble signature inside  
particle detectors could be produced if the millicharge is very small. In order to search the invisible MCPs,
one should rely on the production associated with additional final state particle(s). Such as a photon radiated from the  initial state electron or positron, which has been studied in Refs.  \cite{Liu:2018jdi,Liu:2019ogn,Liang:2019zkb,Gorbunov:2022dgw} via the mono-photon searches. A direct observation for non-relativistic MCPs using energy deposits inside a tracker at electron colliders is investigated in Ref. \cite{Gorbunov:2022bzi}. In this work, we plan to consider a new signature at electron colliders, i.e.,  a light meson $\pi^0$ radiated from the mediated photon in the MCP pair production process $e^+e^-\to\chi\bar\chi$ at electron-positron colliders, which exhibits the mono-$\pi^0$ signature. 
This channel may be an important complement to the mono-photon channel, which will be discussed for the first time.

We will focus on the search for the mono-$\pi^0$ signature at future Super Tau Charm Facility (STCF). The STCF is a projected electron-positron collider operating in the range of center of-mass energies from 2.0 to 7.0 GeV with the peak luminosity of about $10^{35}$  cm$^{-2}$s$^{-1}$ \cite{Achasov:2023gey}. In the following, we will study the MCPs production associated with $\pi^0$ at STCF in Sec. II and the corresponding SM background in Sec. III.
The detector simulation will be discussed in Sec. IV. The numerical results are presented in Sec. V. Finally, we  summarize our findings in Sec. VI.

\section{Millicharged particles production associated with $\pi^0$ at electron colliders} 

\begin{figure}[htbp]
	\vspace{0.2cm}
	\centering
	\includegraphics[scale=1]{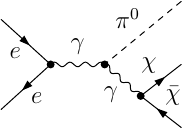}
	\caption{Feynman diagram for the process \eeccp. }
	\label{fig:eecca}
\end{figure}

The Feynman diagram for the single $\pi^0$ production 
associated with millicharged particles \eeccp, 
is shown in Fig.\ref{fig:eecca}. 
The effective interaction Lagrangian for the $\pi^0\gamma\gamma$ coupling can be taken as \cite{Brodsky:1971ud}
\be
{\cal L}_{\pi^0\gamma\gamma}= \frac{g}{2!}\epsilon^{\mu\nu\rho\sigma}F_{\mu\nu}F_{\kappa\lambda}\pi^0,
\ee
where the coupling value $g$ has the relation with the $\pi^0$ decay width $\Gamma_{\pi^0\to\gamma\gamma}$,
\be
g^2=\frac{4\pi \Gamma_{\pi^0\to\gamma\gamma}}{m^3_{\pi^0}}.
\ee
The completely antisymmetric tensor $\epsilon_{\mu\nu\rho\sigma}$ is defined by $\epsilon_{0123}=-\epsilon^{0123}=1$.
The decay width is related to the pion decay constant $f_\pi$ by
\be
\Gamma_{\pi^0\to\gamma\gamma}=\frac{\alpha^2 m^3_{\pi^0}}{64 \pi^3 f_\pi^2}. 
\ee
In this work, we take the mass of the neutral pion $m_{\pi^0}=134.9768$ MeV \cite{ParticleDataGroup:2020ssz}, the 
pion decay constant $f_\pi =0.092388$ GeV \cite{Czyz:2017veo} and the fine structure constant $\alpha=1/137.036$.
From the Lagrangian, one can derive the $\pi^0\gamma^*\gamma^*$ vertex Feynman rule,
$ -4 i  g\epsilon_{\mu\nu\rho\sigma}p^\rho q^\sigma F_{\pi^0\gamma^*\gamma^*}(p^2,q^2)$ \cite{Budassi:2022kqs}, where $p^\rho$ and $q^\sigma$ are the four-momenta of the photons.
The $\pi^0\to\gamma^*\gamma^*$ transition form factor, $F_{\pi^0\gamma^*\gamma^*}\left(p^2,q^2\right)$, dependent on the photon virtualities $p^2$ and $q^2$,  is normalized to $F_{\pi^0\gamma^*\gamma^*}(0,0)=1$.

The corresponding tree-level matrix element for the reaction 
$e^+(p_1) e^-(p_2) \to \chi(q_1) \bar\chi(q_2) \pi^0(k) $,
reads
\bea
{\cal M} &=& \frac{4 i \alpha^2}{f_\pi^0} F_{\pi^0\gamma^*\gamma^*}\left((p_1+p_2)^2,(q_1+q_2)^2\right)\epsilon_{\mu\nu\rho\sigma} \nonumber \\
&\times&\frac{1}{(p_1+p_2)^2(q_1+q_2)^2}(p_1+p_2)^\rho (q_1+q_2)^\sigma \nonumber \\
&\times&\left(\bar v(p_1)\gamma^\mu u(p_2)\right)\left(\bar u(q_2)\gamma^\mu v(q_1)\right),
\label{eq:me-pi0}
\eea
which is similar as the $s-$channel for the process $e^+ e^-\to e^+ e^- \pi^0$ \cite{Czyz:2010sp}.

\section{The SM background}

In practice, the pseduscalar meson $\pi^0$ is reconstructed in the mode $\pi^0\to\gamma\gamma$.
Thus, for the  mono-$\pi^0$ signature at electron colliders, the SM background can arise from  $\pi^0$ direct production 
and the di-photon production with their invariant mass ($M_{\gamma\gamma}$) close to the nominal $\pi^0$ mass.

The SM background can also be divided into two categories according to the final state particles: irreducible background and reducible background.
The irreducible monophoton background is from  neutrino production in the  process $e^+ e^- \to \nu_\ell \bar \nu_\ell \gamma\gamma$ of reconstructing photon pair to $\pi^0$, where
$\nu_\ell = \nu_e, \nu_\mu, \nu_\tau$ are the three standard model 
neutrinos.

The reducible background occurs when $\pi^0$ or photon pair is generated in the final state, along with other visible particles. However, these particles are not detected due to the limitations of the detector acceptance.
The reducible SM backgrounds mainly come from the $e^+e^-\to e^+e^-\pi^0$ 
and $e^+e^-\to e^+e^-\gamma\gamma$ 
processes, where in the final state only $\pi^0$ or $\gamma\gamma$ pair with $M_{\gamma\gamma}\sim m_{\pi^0}$
can be detected in the detectors, and the final state electrons
are undetected because of the limitations of the detectors.
{It is noted that the process $e^+e^-\to \pi^0\gamma$ may contribute to the reducible background, if the associated single photon is emitted along the beam line.
	However, in this case, the event can be easily vetoed, since the reconstructed $\pi^0$ would also be shown as traveling along the beam direction, as it is a $2\to2$ scattering process at a beam-symmetric electron-positron collider (STCF). }

\section{Detector simulation} 
Candidate neutral pions are reconstructed in their diphoton decays.
It is required that the two photons generated by $\pi^0$ decay are both within the effective acceptance of electromagnetic calorimeter (EMC), and then to reconstruct the  $\pi^0$ mass.
The calorimeter at STCF is composed of a barrel part and an endcap cover. The barrel part covers polar angles from $33.85^\circ$
to 146.15$^\circ$. The endcap covers polar angles from 19.18$^\circ$ to 33.64$^\circ$ and 156.15$^\circ$ to 160.82$^\circ$ \cite{Achasov:2023gey}.
The energy coverage of photon detection is from 25 MeV to 3.5 GeV.
Thus, photon candidates to reconstruct  $\pi^0$  are required to 
meet the cut of $E_\gamma >25$ MeV and $|\cos\theta_\gamma|<0.94$ in our analysis.
The invariant mass of the $\gamma \gamma$ pair selected to reconstruct $\pi^0$ candidates needs to be within
the asymmetric intervals [0.115, 0.150] GeV ($0.115 \GeV \leq M_{\gamma\gamma}\leq0.150\GeV)$ \cite{Ablikim:2018fky}.
We will collectively refer to the above cuts for the final photons as the  ``{\it basic cuts}" hereafter.

For the reducible background $e^+e^-\to e^+e^-\pi^0$, 
the final state electron and positron emit along the beam directions with $|\cos\theta|>0.94$ which are undetected beyond the  boundary of the EMC at STCF.
We define the polar angle $\theta_b$ corresponding to the boundary of the EMC with $|\cos\theta_b|=0.94$.
When the electron and positron emit along different directions with $\theta_{e^\pm}=\theta_b$ and have transverse momenta $p_T^{e^\pm}$ opposite to the $\pi^0$ transverse momenta $p_T^{\pi^0}$,  
the maximum energy of the $\pi^0$ meson $E_{\pi^0}^{\rm max}$ in the reducible background 
$e^+e^-\to e^+e^-\pi^0$ occurs for certain polar angle $\theta_{\pi^0}$.
We can obtain $\sqrt{E_{\pi^0}^2-m_{\pi^0}^2}\sin\theta_{\pi^0}=(\sqrt{s}-E_{\pi^0})\sin\theta_b$ by using momentum conservation in the transverse
direction and energy conservation, which leads to the maximum $\pi^0$ energy as a function of its polar angle 

\be
E_{\pi^0}^{\rm max} = \frac{ \sqrt{s}-\sqrt{ s - \left(1-\left(\frac{\sin\theta_{\pi^0}}{\sin\theta_b}\right)^2\right)\left(s+m_{\pi^0}^2 \left(\frac{\sin\theta_{\pi^0}}{\sin\theta_b}\right)^2\right)}}{1-\left(\frac{\sin\theta_{\pi^0}}{\sin\theta_b}\right)^2}.
\ee
In the following, besides the ``{\it basic cuts}" for the photon, we also apply the additional cut $E_{\pi^0}>E_{\pi^0}^{\rm max} $ on the final reconstructed $\pi^0$, which will be labeled as ``{\it advanced cuts}",  in our analysis to suppress
the events arising from the   reducible backgrounds, such as   $e^+e^-\to e^+e^-\pi^0$ as well as $e^+e^-\to e^+e^-\gamma\gamma$.

We use the {\sc FeynArts} \cite{Hahn:2000kx} and {\sc FormCalc} \cite{Hahn:1998yk} packages 
to simulate the background from $e^+ e^- \to \nu_\ell \bar \nu_\ell \gamma\gamma$ and  $e^+ e^- \to e^+ e^- \gamma\gamma$ processes.
Regarding to $e^+ e^- \to e^+ e^- \gamma\gamma$ processes, we take the cut $|\cos\theta|>0.94$ for the final state electron and positron, which is beyond the coverage of the EMC.
The events for the reducible background $e^+e^-\to e^+e^-\pi^0$ is generated by the {\sc  EKHARA 3.0}  package \cite{Czyz:2018jpp}.
For the signal process \eeccp, with the help of {\sc FeynCalc} package \cite{Shtabovenko:2020gxv},
we can obtain the squared matrix element in Eq. (\ref{eq:m2}) and 
import it to the {\sc FormCalc} \cite{Hahn:1998yk} package to numerically evaluate
the cross section.

As a cross check of our Monte Carlo results, we
produced the results  on the
total cross section of the process $e^+e^-\to \chi\bar\chi\pi^0$ by setting $m_\chi=m_e$ and $\varepsilon=1$ with the same input parameter, 
and  obtained excellent agreement between our prediction and the $s$-channel  cross section of the process $e^+e^-\to e^+e^-\pi^0$ obtained with  {\sc  EKHARA 3.0} package  \cite{Czyz:2018jpp}.

In our following calculation, the form factor $F_{\pi^0\gamma^*\gamma^*}$  has been calculated according to the resonance chiral symmetric model
with SU(3) breaking of Ref.  \cite{Czyz:2017veo}, which is fitted to the widest data set, both in the space-like and time-like regions
and  recommended in {\sc  EKHARA 3.0} package\cite{Czyz:2018jpp}.

\section{Numerical results for mono-$\pi^0$ production}

\begin{figure*}
	\begin{center}
		\includegraphics[angle=0,width=3.2in,height=2.4in]{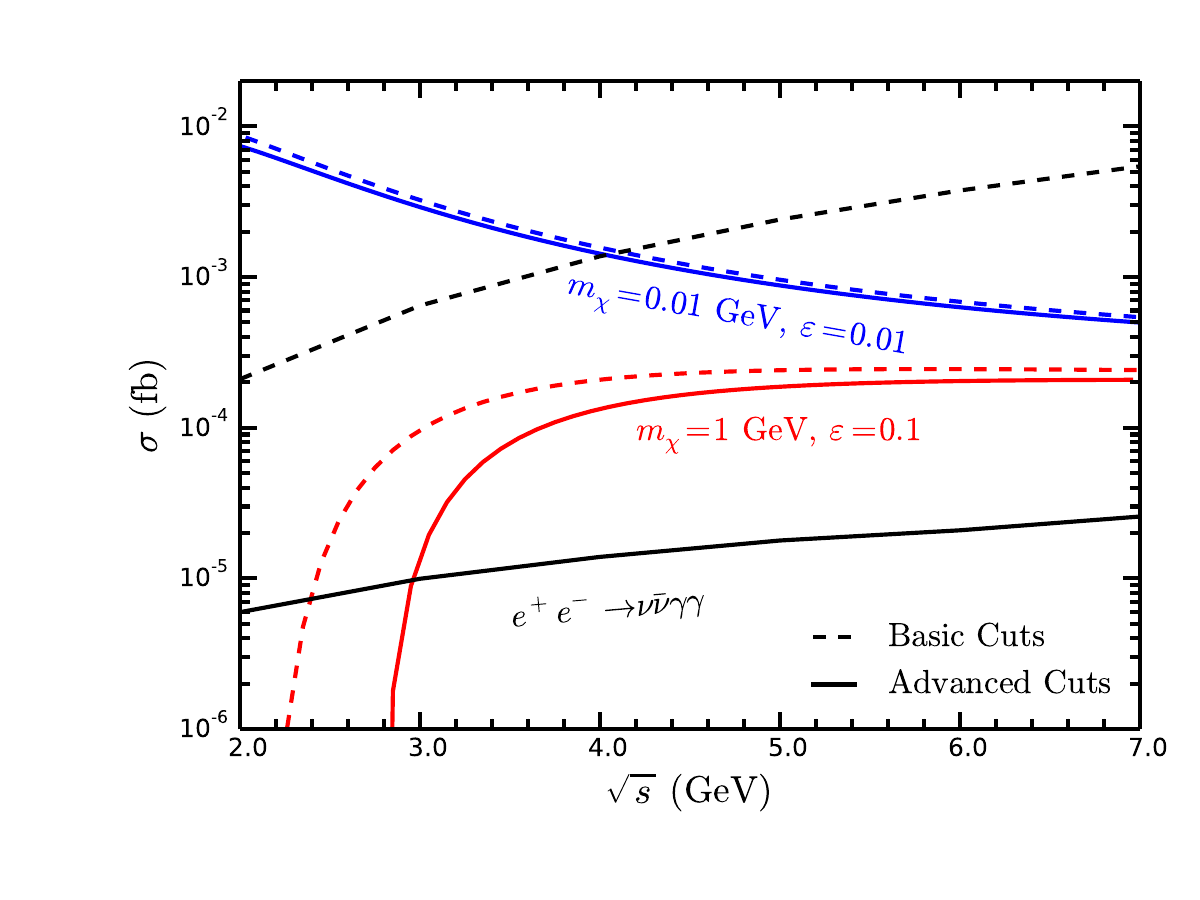}
		\includegraphics[angle=0,width=3.2in,height=2.4in]{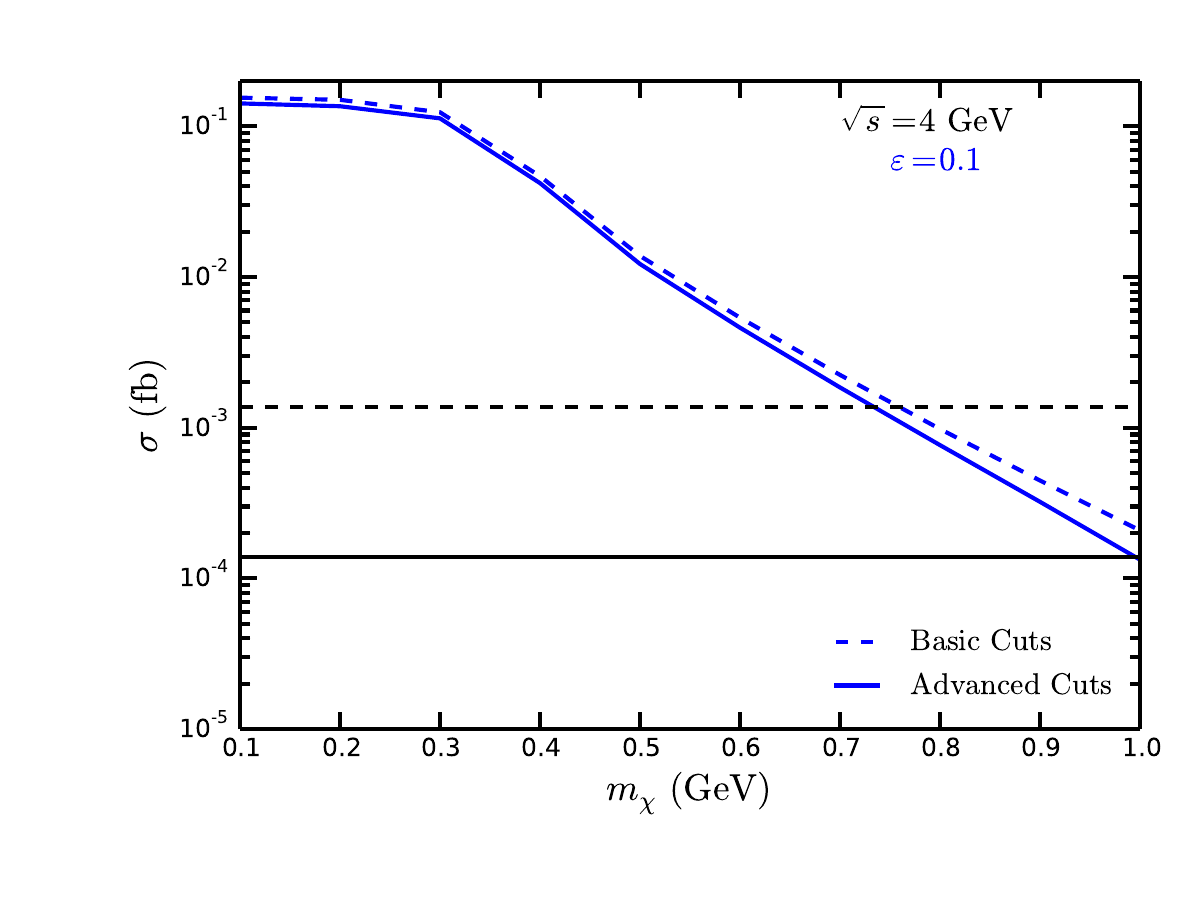}
		\caption{The mono-$\pi^0$ production rates from the irreducible SM background $e^+e^-\to \nu\bar{\nu}\gamma\gamma$ and from millicharged particles as a function of $\sqrt s$ in 
			the STCF energy range, $2.0 \,{\rm GeV} \le \sqrt s \le 7.0 \,{\rm GeV}$ (left) and of the millicharged particle mass $m_\chi$ (right)  after  ``{\it basic cuts}" (dashed lines) or ``{\it advanced cuts}" (solid lines) for the final photon and reconstructed  $\pi^0$, respectively. 
		}
		\label{fig:xsec}
	\end{center}
\end{figure*}

In Fig.\ref{fig:xsec}, we present the cross sections $\sigma$ for the signal and irreducible backgrounds at future STCF as the function of running energy $\sqrt{s}$ (left) and  the mass of millicharged particle $m_\chi$ (right) after the above mentioned ``{\it basic cuts}" or ``{\it advanced cuts}" for the final photon and reconstructed  $\pi^0$, respectively. 

From the figure, one sees that the production rates for light mass millicharged particle ($m_\chi=0.01 \GeV$)  associated with single $\pi^0$ decrease with the increment of colliding energy, while the cross section due to the SM irreducible processes always increases. As a consequence, the signal to background ratio increases when the colliding energy decreases.
For the heavier millicharged particle ($m_\chi=1 \GeV$), the dependence of production rates on $\sqrt{s}$ is completely different.
Due to the suppression of kinematics, it's hard to produce heavier millicharged particle with lower colliding energy.
With the increment of $\sqrt{s}$, the cross section for the signal exhibits a rapid growth trend. However, 
when $\sqrt{s}\gtrsim 4\GeV$,  its growth rate significantly slows down and gradually begins to stabilize.

For the graph on $m_\chi-\sigma$ plane with $\sqrt{s}=4\GeV$ in Fig.\ref{fig:xsec} on the right, we observe that for the light millicharged particle with $m_\chi\lesssim 0.3\GeV$, the production cross section diminishes gradually with the increase in $m_\chi$, exhibiting near-independence from $m_\chi$. Nevertheless, as the mass continues to increase, the cross section experiences a rapid decline.

After adopting  the ``{\it advanced cuts}", one can find that the irreducible background $e^+e^-\to\nu\bar\nu\gamma\gamma$ can be reduced by over one order of magnitude, even  reaching two orders of magnitude. Conversely, the signal is less affected. Assuming total luminosity of 20 ab$^{-1}$ collected in the future, the number of events due to the irreducible background will be less than one, which could be barely background-free with  the ``{\it advanced cuts}" since the reducible background is removed. 
Thus the mono-$\pi^0$ could be a clean signature to search the new physics beyond SM at future STCF.


\begin{figure*}
	\begin{center}
		\includegraphics[scale=0.7]{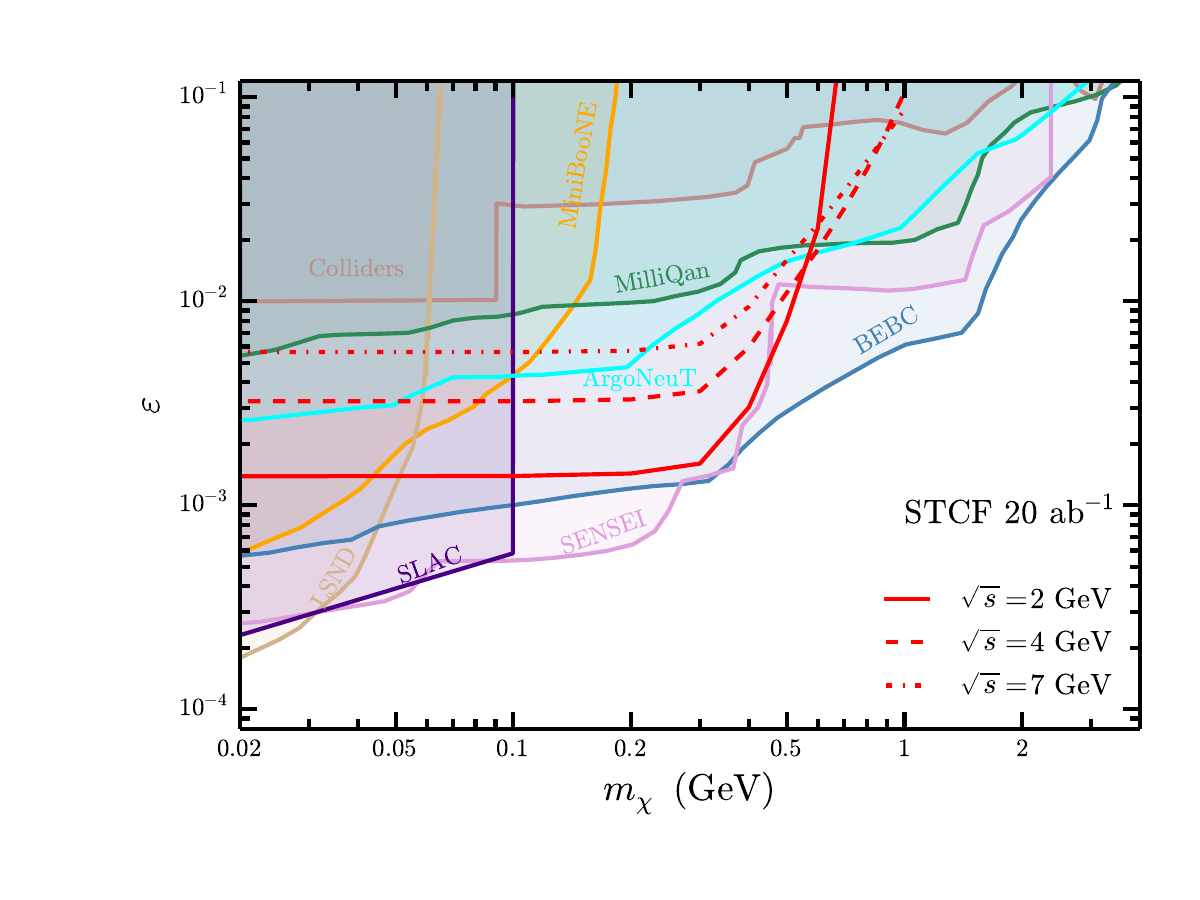}
		\caption{The expected 95\% C.L. exclusion limits on MCPs at future STCF via mono-$\pi^0$ searches with integrated luminosity of 20 ab$^{-1}$ running at $\sqrt{s}= 2$ GeV (red solid line), 4 GeV (red dashed line) and 7 GeV (red dot-dashed line), respectively. The current constraints are shown as  shaded regions from SLAC \cite{Prinz:1998ua}, LSND \cite{Magill:2018tbb}, Colliders \cite{Davidson:2000hf,Liang:2019zkb}, MiniBooNE \cite{Magill:2018tbb}, MiiliQan \cite{Ball:2020dnx}, ArgoNeuT \cite{ArgoNeuT:2019ckq}, BEBC\cite{Marocco:2020dqu}, SENSI \cite{SENSEI:2023gie}. }
		\label{fig:result}
	\end{center}
\end{figure*}

In Fig.\ (\ref{fig:result}), we show the  95\% confidence level (C.L.) exclusion upper limits on millicharge  
$\varepsilon$ for various masses, by assuming collected integrated luminosity of 20 ab$^{-1}$ with $\sqrt{s}= 2$ GeV, 4 GeV and 7 GeV,
respectively. The 95\% C.L. upper limits are obtained by solving $\chi^2(\varepsilon_{95})-\chi^2(0)=2.71$ with the definition of $\chi^2\equiv S^2/(S+B)$, where $S(B)$ is the number of events in the signal (background) processes.  The existing limits from other experiments are also shown.
One can see that, via the mono-$\pi^0$ searches at future STCF, for the light millicharged particle ($m_\chi\lesssim 0.5$ GeV), with the same luminosity higher colliding energy can provide stronger sensitivity. Specifically speaking, STCF running at $\sqrt{s}= 2$ GeV with assuming collected integrated luminosity of 20 ab$^{-1}$ can probe the millicharge down to about $ 1.6 \times 10^{-3}$ for the $m_\chi \lesssim 0.2$ GeV case. Via the mono-$\pi^0$ searches, the $\sqrt{s}= 2$ GeV STCF with luminosity of 20 ab$^{-1}$ can improve  previous bounds from ArgoNeuT \cite{ArgoNeuT:2019ckq}  as much as a factor of 8 in the mass range from 100 MeV to 400 MeV, but these limits are fully excluded by the latest derivation from the  past BEBC WA66 beam dump experiment \cite{BEBCWA66:1986err, Marocco:2020dqu} and a new SENSEI experiment \cite{SENSEI:2023gie}.

\section{Summary}

In this work, we have proposed a new search for {millicharged} particles 
via the mono-$\pi^0$ signature at future STCF for the first time. We present the analytical matrix element and provide  numerical codes to calculate the production rates for the MCP pair production with a light meson $\pi^0$. 
We find that by using  the assumed integrated luminosity of 20 ab$^{-1}$, the STCF running at $\sqrt{s} =2$ GeV or 4 GeV can provide leading constraints 
compared with existing results from ArgoNeuT \cite{ArgoNeuT:2019ckq}  in the  mass range, 
$0.1~\text{GeV} \lesssim m_\chi \lesssim 0.6~\text{GeV}$, but excluded  by the latest derivation from the  past BEBC WA66 beam dump experiment \cite{BEBCWA66:1986err, Marocco:2020dqu}.
We also systematically analyzed the irreducible and 
reducible SM backgrounds for the 
 mono-$\pi^0$ signature at the future STCF 
which was lacking in the literature to our 
knowledge. We find that mono-$\pi^0$ searches is barely background-free, which could be a clean channel to search for the invisible particles, such as dark matter.

\acknowledgments
We thank Zeren Simon Wang for helpful correspondence and discussions. 
This work is supported by the National Natural Science Foundation of China under grant No. 12475106 and the Fundamental Research Funds for the Central Universities (JZ2023HGTB0222).


\begin{widetext}
	
\appendix 
\section{The analytic expression of the squared matrix element for the signal process \eeccp}
\label{appendix: M2}
The squared matrix element for the signal process for the  $e^+(p_1) e^-(p_2) \to \chi(q_1) \bar\chi(q_2) \pi^0(k) $ in the millicharged particle production associated with $\pi^0$ can be expressed as,
\begin{equation}
\begin{aligned}
|{\cal M}|^2 &= \big(\frac{4  \alpha^2}{f_\pi^0} F_{\pi^0\gamma^*\gamma^*}\big)^2
2\Big[2 m_e^2 \Big(2 m_{\chi }^2 \big(2 s_1 \left(s_2-t_1-t_2\right) 
- 2 s_2 \left(t_1+t_2\right)+s_1^2+s_2^2-4 s s_{34}  + t_1^2+t_2^2+2 t_1 t_2\big)\\
&+s_{34} t_1^2+s_{34}t_2^2 
- 2 s s_{34} t_1-2 s_2 s_{34} t_1-2 s s_{34} t_2+2 s_2 s_{34} t_2-2 s_2 t_1 t_2 -2 s_{34} t_1t_2 \\
&+2 s_1 \big(-s_2 \left(s_{34}+t_1+t_2\right)+s_{34} t_1-s_{34} t_2+s_2^2-s s_{34}-t_1
t_2\big)+s_2^2 s_{34}-2 s s_2 s_{34}\\ 
&+s_1^2 \left(2 s_2+s_{34}\right)+2 t_1 t_2^2+2 t_1^2t_2\Big) \\
&-2 m_e^4 \Big(2 s_1 \left(s_2-t_1-t_2\right)-2 s_2 \left(t_1+t_2\right)  +s_1^2+s_2^2-4s s_{34}+t_1^2+t_2^2+2 t_1 t_2\Big) \\ 
& -2 m_{\chi }^4 \Big(2 s_1 \left(s_2-t_1-t_2\right)-2 s_2
\left(t_1+t_2\right)+s_1^2+s_2^2-4 s s_{34}+t_1^2+t_2^2+2 t_1 t_2\Big)\\
&+2 m_{\chi }^2 \Big(s
\big(-2 s_1 \left(s_2+s_{34}+t_1-t_2\right) -2 s_{34} t_1-2 s_{34} t_2-2 s_2
\left(s_{34}-t_1+t_2\right) \\
&+s_1^2+s_2^2+t_1^2+t_2^2-2 t_1 t_2\big) +2 \big(s_1 \left(-s_2
\left(t_1+t_2\right)+s_2^2-t_1 t_2\right) +t_1 t_2 \left(-s_2+t_1+t_2\right)+s_2
s_1^2\big)\Big) \\
&-2 s^2 s_{34}^2+s s_{34} t_1^2+s s_{34} t_2^2+4 s_1 s_2 t_1 t_2+2 s s_{34}
t_1 t_2-2 s_1^2 s_2^2+s s_1^2 s_{34}+s s_2^2 s_{34}+2 s s_1 s_2 s_{34}-2 t_1^2 t_2^2\Big],
\label{eq:m2}
\end{aligned}
\end{equation}
where the six kinematic invariants are $s=(p_1+p_2)^2$, $s_{34}=(q_1+q_2)^2$, $s_1=(p_1+q_2)^2$, $s_2=(p_2+q_1)^2$, $t_1=(p_1-q_1)^2$, $t_2=(p_2-q_2)^2$.

\end{widetext}

\normalem


\end{document}